\documentclass{elsart5p}
\usepackage[english]{babel}
\usepackage{graphicx}
\usepackage{graphicx}
\usepackage{amsmath}
\usepackage{subfigure}
\usepackage{amssymb}
\usepackage[dvips]{color}
\usepackage{epsf}
\begin{document}
\begin{frontmatter}
\title{Molecular and all solid DFT studies of the magnetic and chemical bonding properties within KM[Cr(CN)$_6$] (M = V, Ni) complexes.}
\author[AA]{L. Kabalan},
 \author[AA]{S.F. Matar\corauthref{S.F. Matar}}, 
\author[AA]{C. Desplanches}, 
\author[AA]{J.F. L\'etard},
\author[BB]{M. Zakhour}.
\address[AA]{Institut de Chimie de la Mati\`ere Condens\'ee de Bordeaux, CNRS,  Universit\'e Bordeaux 1. Pessac, 87
Avenue du Dr. Albert Schweitzer, F-33608 Pessac Cedex, France}  
\address[BB]{Laboratoire de Chimie Physique des Mat\'eriaux, Facult\'e des Sciences, Universit\'e Libanaise, Fanar-Beyrouth, Lebanon.}

\corauth[S.F. Matar]{Corresponding author. Tel: +33 540002690. Fax: +33 540002761. email: matar$@$icmcb-bordeaux.cnrs.fr}
\begin{abstract}
 A study at both the molecular and extended solid level in the framework DFT is carried out for KM[Cr(CN)$_6$] ( M = V, Ni). From molecular calculations, the exchange parameters J are obtained, pointing to the expected magnetic ground states, i.e., antiferromagnetic for M = V with J = -296.5 cm$^{-1}$ and ferromagnetic for M = Ni with J = +40.5 cm$^{-1}$. From solid state computations the same ground states and J magnitudes are confirmed from energy differences. Furthermore an analysis of the site projected density of states and of the chemical bonding is developed in which the cyanide ion linkage is analyzed addressing some isomerism aspects. 
\end{abstract}
\begin{keyword}
 DFT \sep exchange parameters\sep ferromagnets \sep antiferromagnets \sep ASW.
\PACS {71.15.Mb, 71.15.Nc, 71.20-b, 75.10.Lp, 74.25.Ha}   
\end{keyword}
\end{frontmatter}
\section{\bf Introduction}
\begin{figure}[htbp]
\begin{center}
\includegraphics[height= 0.15\textheight]{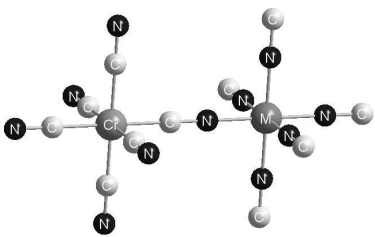}
\caption {Sketch of the KM[Cr(CN)$_6$] structure. M stands for the other transition metal ion, V or Ni.}
\label{fig1}
\end{center}
\end{figure}

Prussian blue KFe$^{III}$[Fe$^{II}$(CN)$_6$], is a historically known chemical compound \cite{Anonymous}. The presence of two types of cationic transition metals (TM) as shown in fig. \ref{fig1} allows to extend the family to several systems by substitutions not only of TM but of K as well by other monovalent alcaline cations such as Cs  \cite{Dunkar}. Although the crystal structure is face centered cubic with high symmetry (space group F-43m, N$^o$216), the elucidation of the Prussian blue structure was the object of intense research work due to the crystallographic problems \cite{Kahn2}. Nowadays, the study of magnetism in such systems is a rich field of investigation and many magnetic materials have been synthesized and characterized. 
Furthermore, due to the high symmetry of the structure, molecular field theory can be easily applied to these types of systems \cite{Kahn}. Therefore, Prussian blue analogues are the obvious choice for the design of molecular magnets, and many fascinating and exciting phenomena have been reported in the literature. Examples of such studies include high ordering-temperature magnets \cite{Verdaguer, Gadet, Ferlay}, photoinduced magnetism \cite{Sato}, electrochemically tunable magnets \cite{Sato2}, and magnets that experience two compensation temperatures \cite{Ohkoshi}. Most of these examples have been designed taking advantage of the properties of this family of compounds. The solids of Prussian blue analogues are readily prepared from cyanometallate blocks and transition metal cations. 
\newline
 The first quantum theoretical studies were done on the molecular systems using the extended-H\"uckel method for dinuclear models \cite{Verdaguer, Verdaguer2}. They applied the Kahn-Briat model \cite{Kahn1} and attributed the exchange interaction to the overlap between magnetic orbitals through the $\pi$ system. Many other studies on the exchange interactions were carried out to evaluate the magnitude of J coupling parameter allowing an assignment of the type of magnetic order from molecular calculations \cite{Ruiz, Nishino}. For the solid state periodic structures, the work of Eyert et al. \cite{Eyert} on two well-characterized Prussian blue analogues of formula CsM[Cr(CN)$_6$] (M = Mn, Ni)  can be noted within density functional theory (DFT) \cite{Honenberg}. 
 \newline
In this work, we carry out a study at both the molecular and extended solid levels within DFT for KM[Cr(CN)$_6$] (M = V, Ni). Exchange parameters  J, are obtained from molecular calculations and confronted with J values from energy differences within solid state computations assuming different magnetic configurations, namely non-magnetic (NM), ferromagnetic (FM) and antiferromagnetic (AFM) configurations. Special attention will be devoted to an analysis of the site projected density of states and a study to the chemical bonding is developed in which the cyanide ion linkage is analysed addressing isomerism aspects. 

This double approach at the molecular and solid state levels has already allowed us to address the low spin and high spin properties of the spin cross-over complex [Fe(btz)$_2$(NCS)$_2$] \cite {Kabalan}.

\section{\bf Magnetic orders}
Depending upon the response of a material to an applied field, we can describe it as either a diamagnetic or paramagnetic sample. Paramagnetic materials may be further classifield as ferromagnets, ferrimagnets and antiferromagnets.
Table I sketches out the ferromagnetic superexchange interactions through the cyanide bridge between d orbitals of Cr and M  with O$_h$ environment. This diagram involves a difference of electron occupation of e$_g$ which can lead to ferromagnetic or antiferromagnetic interactions. When the interaction is between t$_{2g}$ - e$_g$ which is the case of Cr-Ni, this leads to a ferromagnetic interaction due to the electronic structure of the cyanide bridge and special role of nitrogen, while t$_{2g}$Cr - t$_{2g}$V leads to antiferromagnetic interactions \cite {Matar}. 

Ferromagnets, behave as paramagnets at high temperatures, but below a critical  temperature T$_C$ (Curie temperature), they show magnetic ordering. In a mean field approximation, it can be demonstrated that: 
\begin{eqnarray}
\label{TC}
T_C =  \frac{ {\sqrt{Z_M Z_{M'}}} |J|  \sqrt {x  S_M (S_M +1) S_{M'} (S_{M'}+1)}}{3k_B}
\end{eqnarray}
where S$_M$ and S$_{M}^{'}$ are the local spins on centers M and M$^{'}$, Z$_M$ and Z$_{M}^{'}$  the number of nearest neighbors of each type of metal atom and x corresponds to the stoichiometry of the M'[M(CN)$_6$]$_x$ compound. J is the exchange coupling between the two TM. It indicates the type of communication between the two spin centers. A positive value of J represents a ferromagnetic coupling, while negative J describes an antiferromagnetic interaction. 
The J values are obtained by using equation: 
\begin{eqnarray}
\label{energy}
E_{HS} - E_{LS}  =  - (2S_1S_2 + S_2)J
\end{eqnarray}
where S$_1$ and S$_2$ are the total spins of the paramagnetic centers and S$_1$ $>$ S$_2$ has been assumed for heterodinuclear complexes. E$_{HS}$ is the energy of highest spin state and E$_{LS}$ that of the sate having the two metals with opposite spins.
\newline 
Over the last decades, the Prussian blue analogues A$_l$M${_{m }^ {'} }$[M(CN)$_6$](H$_2$O), have received attention as molecule-based magnetic materials. Effective exchange interactions between M and M$ ^{'}$ via the CN$^-$ anion (note that the negative charge is carried by carbon strong leading to high T$_C$ values. Verdaguer and coworkers have synthesized compounds with T$_C$ value of 90, 240 and 315 K, with M$^{'}$  = Ni(II) and M = Cr(III) \cite{Gadet, Mallah}, M$^{'}$ = Cr(II) and M = Cr(III) \cite{Mallah2}, and M$^{'}$ = V(II) and V(III), M = Cr(III) \cite{Ferlay} respectively. 

\section{\bf Computational Details}
\subsection{Calculations at the molecular level}
Although the use of Hartree-Fock (HF) approach has been shown to provide a good description of the molecular orbital and chemical bonding properties mainly in organic chemistry, it becomes well established that calling for the density functional theory (DFT) framework \cite{Honenberg} brings far more accurate results regarding the energetics and related properties. This is because the compulsory exchange and correlation (XC) effects are equally treated, albeit at a local level, within DFT while only exchange is well accounted for in HF although in a better way (exact exchange \cite{Becke}) than in DFT. Taking the best out of each one of the two approaches led to improvements in ab initio molecular calculations with the so called "hybrid functionals". They consist of mixing exact HF exchange, ex. following Becke \cite{Becke} and DFT based correlation, ex. following Lee, Yang and Parr, i.e., the so-called LYP correlation \cite{Lee}, with proportions that help to reproduce molecular properties of several systems. The calculations have been performed using $Gaussian03$ package \cite{G03}. Owing to the complexity of the electronic structure of the studied compounds, we have employed the {\it JAGUAR} code \cite{Jaguar} to generate an appropriate initial wavefunction  set since it provides a better control of the local spin and multiplicities of the atoms. To maintain the electronic structure during the SCF procedure the quadratic convergence option must be used. In all calculations, the hybrid functional B3LYP  and the LanL2DZ basis sets were used. The latter  includes double-${\zeta}$ with the Los Alamos effective core potential for Cr, V, Ni and the Dunning-Huzinaga all-electron double-${\zeta}$ basis set with polarization functions for the H, C, and N atoms (${\zeta}$ is the exponent in the Gaussian type orbitals GTO) ~\cite{Dunning, Hay}. The following bond lengths used for the calculations, were for the KV[Cr(CN)$_6$] complex: d$_{Cr-C}$ = 2.01 \AA, d$_{V-N}$ = 2.01\AA~ and d$_{C-N}$ = 1.24 \AA. For the second complex the distances were : 2.03, 2.04 and 1.24 for d$_{Cr-C}$, d$_{V-N}$ and d$_{C-N}$ respectively.

\subsection{Solid state computations}
In as far as no structural determinations were available for both systems, trends in cell volumes and ground state crystal structure were desirable in the first place. The equilibrium structures were obtained from a full geometry relaxation from CsNi[Cr(CN)$_6$] structural setup \cite {Eyert} and using a pseudo-potential approach within the VASP package \cite{Kresse}. Ultra-soft Vanderbilt pseudo-potentials (US-PP) built within the local density approximation (LDA) scheme \cite{lda} were used. The Brillouin-zone integrals were approximated using a special $k$-point sampling. These values were then used in the all electron augmented spherical wave method (ASW)  \cite{ASW}. Besides its use of DFT, this method is based on the atomic sphere approximation (ASA), a special form of muffin-tin approximation which consists in dividing the cell volume into atomic spheres whose total volume is equal to the cell volume. The calculation being carried out in the atomic spheres space, empty spheres (pseudo-atoms) need to be introduced -without symmetry breaking- in open (low compactness) structures such as that of the systems studied here. Further, empty spheres allow for the iono-covalent characteristics of the system to be accounted for by receiving charges from actual atomic species. The use of this method in molecular systems was formerly validated by Eyert et al. \cite{Eyert}. The Brillouin-zone sampling was done using an increased number of k points ranging from 126 to 840 points within the irreductible wedge. In this way we were able to ensure convergence of the results with respect to the fineness of the k-space grid.

In the context of this study, ASW-LDA was used  in order to obtain a description of the electronic structure with the partial, site projected, density of states DOS. Further the chemical interaction is assessed by using the ECOV (covalence bond energy) criterion, which allows to get two-body chemical bond characteristics \cite{Ecov}. Negative, positive and zero ECOV magnitudes  point to bonding, anti-bonding and non-bonding interactions respectively. 
In order to enable for a better understanding of J, the electronic properties as well as the DOS and the ECOV, we start out by discussing  the results for both compounds. The strategy of the ASW calculations was to do three types of computations for the two complexes : Firstly a non-magnetic NM calculation (degenerate spins), then detecting an instability of the NM configuration magnetic calculation, a Ferromagnetic calculation is carried out. Finally, the system is enforced to be anti-ferromagnetic through supercell calculations in order to establish the ground state expected for KV[Cr(CN)$_6$]. 

\section{\bf Results and discussion}
\subsection{Calculation of J}
\begin{figure}[htbp]
\begin{center}
\subfigure[~]{\includegraphics[width=0.8\linewidth]{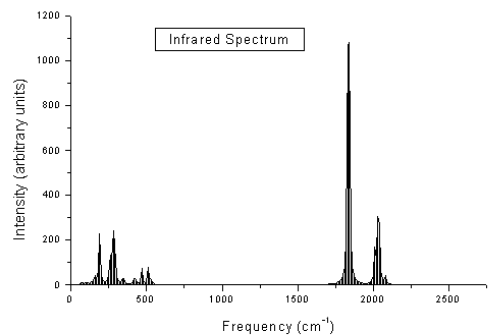}\label{1a}}
\subfigure[~]{\includegraphics[width=0.8\linewidth]{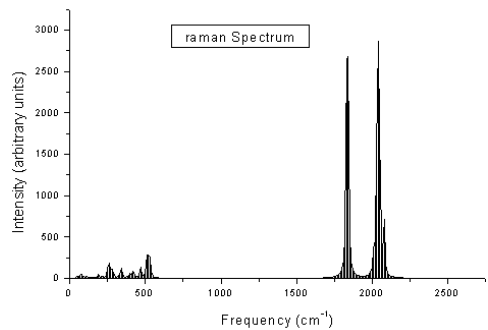}\label{1b}}
\caption{KV[Cr(CN)$_6$]: a) IR and b) Raman Spectrum calculated with  B3LYP hybrid functional and LanL2DZ basis set.}
\label{fig2}
\end{center}
\end{figure}

The three transition metals, chromium, vanadium and nickel, ionize as Cr$^{III}$ (d$^3$), V$^{II}$ (d$^3$) and Ni$^{II}$ (d$^8$) within the complexes KM[Cr(CN)$_6$] under consideration. These formal degrees of ionization are fed into {\it JAGUAR} input in order to calculate the charge densities of the different magnetic alignment configurations, i.e., FM and AFM. Then with unrestricted HF (accounting for spins) calculations the energy differences (eqn. \ref{energy}) allow obtaining J.   
For M =  V system, a very strong antiferromagnetic coupling with a large J = -296 cm$^{-1}$ value was found. This corresponds to the scheme of V$^{II}$-t$_{2g}^3$ $\longleftrightarrow$ Cr$^{III}$-t$_{2g}^3$ antiparallel coupling with O$_h$ environment, in both e$_g$ manifolds are empty (see lower part of Table 1). On the contrary a ferromagnetic coupling with J = + 40.5 cm$^{-1}$ value is found to agree with the Cr$^{III}$-Ni$^{II}$ parallel coupling of spins: t$_{2g}^3$ $\longleftrightarrow$ t$_{2g}^6$e$_g^2$. For these two complexes, Ruiz et al \cite{Ruiz}  have obtained the values of J = -241 cm$^{-1}$ and +22.4 cm$^{-1}$.  From this outcome  our values follow the same trends but the differences can be due to the variation of the distances and the basis set they use, i.e., TZV ( triple ${\zeta}$). T$_C$ calculated via eqn. \ref{TC},  gives a value of 3171 K (for 315 K) for the KV[Cr(CN)$_6$], and 274 K for KNi[Cr(CN)$_6$] (for 90 K) \cite{Ruiz}. This overestimation of the T$_C$ values is a well-known problem.
The electron configuration of M and M' affects the nature and magnitude of the coupling in a predictable way, t$_{2g}^m$-t$_{2g}^n$ pairs giving strong antiferromagnetic interaction (case of Cr-V) while t$_{2g}^m$-t$_{2g}^6$ pairs yielding ferromagnetic coupling (case of Cr-Ni) (See Table 1). This result is in agreement with those obtained by Ruiz \cite{Ruiz} who have proposed these explanations based on the calculations with several transition metals. 
Table 2 regroups the atomic spin densities on Cr, V and Ni calculated with the hybrid functional B3LYP  and the LanL2DZ basis sets. The density of state on Cr varies slightly from ferromagnetic to the antiferromagnetic state.
\newline
In addition to the calculation of J, the infra-red (IR) and Raman theoretical frequency spectra were computed  by using also the hybrid functional B3LYP  and the LanL2DZ basis sets cf. fig. \ref{fig2}. The most important peaks are those at $\nu $ =189.017 cm$^{-1}$ assigned to the V-N stretching, Cr-C elongation appears at 255.447 cm$^{-1}$, the high intensity at 1835.03 cm$^{-1}$ is that of elongation of the C = N between the two metals. The latter is also found in Raman spectrum. The elongation of the other C=N  appear at 2051.27 cm$^{-1}$ as an intense peak.
\newline
The frequency of the electromagnetic wave which induces the vibration of elongation is given by the relation 
\begin{eqnarray}
\label{k}
\nu=\frac{1}{2\pi} \sqrt{\frac{k}{\mu}}
\end{eqnarray} 
where k is the constant of bonding strength (considered here as a spring), proportional to the binding energy and $\mu$ the reduced mass of the two atoms connected by this bond. Feeding the $\nu$ values in eqn. \ref{k} for Cr-C and V-N provides a ratio of force constants k, as  k($\frac{Cr-C}{V-N}$) = 1.62 which means that the strength of Cr-C bond is 1.62 larger than that of V-N bond.
This will be confirmed in the study of the chemical bond in the solid state. 
\subsection{ASW calculations}
KV[Cr(CN)$_6$] and KNi[Cr(CN)$_6$] both crystallize in a face-centered cubic lattice with space group F$\bar {43}$m \cite{Gadet}. In this space group K, M and Cr are found at fixed Wyckoff sites, i.e. respectively at (4c/4d), (4a) and (4b) while the (24f) position at x,0,0 allows for variable x value. The lattice relaxation with VASP US-PP calculations keep the fcc symmetry and give the following values for (24f) positions: 
x$^{V/Ni}_{C}$ = 0.309/ 0.306 and x$^{V/Ni}_{N}$ = 0.192/0.188. A good agreement with literature \cite{Gadet} is found for the C and N x positions for the M = Ni, Cs-based system. The relaxed lattice constants are a$_{KV[Cr(CN)_6]}$ = 10.52 \AA ~ and a$_{KNi[Cr(CN)_6]}$ = 10.64 \AA ~ which come in close magnitudes to other similar systems \cite{Eyert}. 

\subsubsection{Nonmagnetic calculations for KV[Cr(CN)$_6$]}
At self-consistent convergence of the energy and of charges, the charge transfer is found from TM towards C and N as well as to the ES with a larger magnitude. In this hypothetical magnetic configuration the system is found to behave as a metal due to the large density of states at the Fermi level. This is exhibited in the site projected partial densities of state (PDOS) in fig. \ref{fig4}a illustrating the results for KV[Cr(CN)$_6$] member. All energies are relative to the Fermi level E$_F$. In the closely octahedral (O$_h$) crystal field of CN ligand surrounding the TM ions, V and Cr, the d orbitals split into two main manifolds which are labeled as t$_{2g}$ (three-fold) and e$_g$ (two-fold). In the PDOS plots three dominant peaks are identified for Cr at -4 eV for the states which mix with CN ligands, at E$_F$ for non-bonding t$_{2g}$ orbitals and at 3.5 eV, in the conduction band, for empty e$_g$. Vanadium states are characterized similarly with DOS peaks at -6 eV, at E$_F$ for t$_{2g}$ and at 2.5 eV for empty e$_g$. Carbon and nitrogen DOS in the lower energy part of the valence band, i.e. in the [-6,-2 eV] energy range are very similar pointing to the covalent C...N bonding, in fact the cyanide is built through strong interactions (one $\sigma$ and two $\pi$) between carbon and nitrogen orbitals. 

The very high density of states at the Fermi level characterizing V and Cr is a sign of instability for the system in such a NM configuration. This is analyzed through the Stoner mean field theory of band ferromagnetism \cite{Stoner}. If the energy of the system is counted from a NM configuration, it can be expressed as follows: 
\begin{eqnarray}
\label{stoner}
E = \frac{1}{2} \frac{m^2}{n _{EF}}  [1-I.n_{EF}].
\end{eqnarray}
In this equation, m is the magnetic moment, I is the Stoner exchange-correlation integral which is an atomic quantity that can be derived from spin polarized calculations, n$_{EF}$ represents the PDOS value of the element under consideration at the Fermi level in the NM state. Between the square brackets in eqn. \ref{stoner}, the product I.n$_{EF}$ can be considered as criterion of instability of the system: if it is  $ > $1, then the  energy  counted from NM configuration is lowered because 1-I.n$_{EF}$ $<$ 0,  hence the system is stabilized by magnetic exchange.
The product I.n$_{EF}$ in eqn. \ref{stoner} is then considered as a criterion for the stability of the spin system. 
For V, I.n$_{EF}$ = 1.831 and in the case of Cr it is equal to 1.900. From this, the system is unstable in a nonmagnetic configuration.

\begin{figure}[htbp]
\begin{center}
\subfigure[~]{\includegraphics[width=0.80\linewidth]{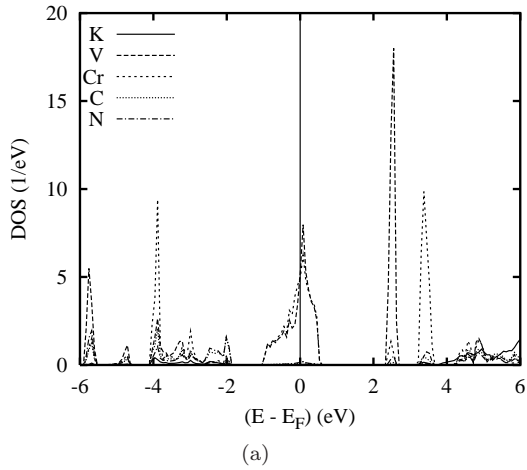}\label{1a}}
\subfigure[~]{\includegraphics[width=0.80\linewidth]{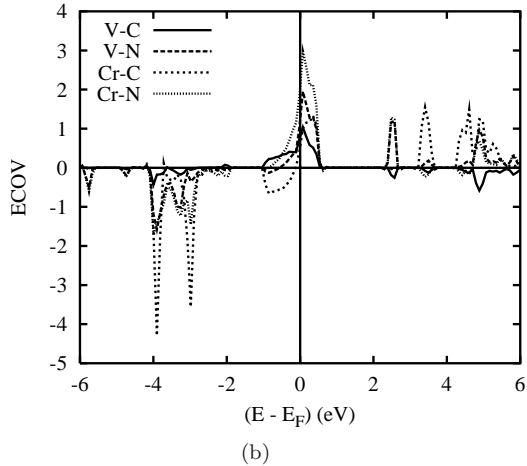}\label{1b}}
\caption{KV[Cr(CN)$_6$] in NM configuration: a) Site projected density of states;  b) Chemical bonding from ECOV criterion for two atom interactions: V-C, V-N, Cr-C and Cr-N.}
\label{fig3}
\end{center}
\end{figure}

\begin{figure}[htbp]
\begin{center}
\subfigure[~]{\includegraphics[width=0.8\linewidth]{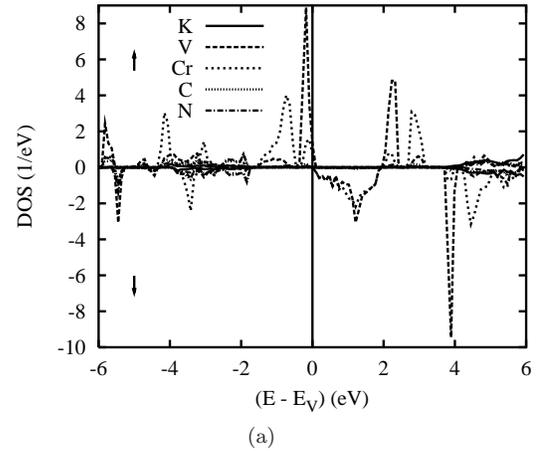}\label{1a}}
\subfigure[~]{\includegraphics[width=0.8\linewidth]{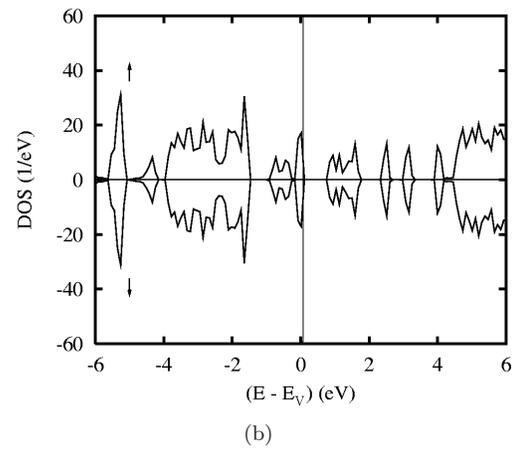}\label{1b}}
\caption{KV[Cr(CN)$_6$] in SP configurations:  a) FM site projected density of states, b) AFM total DOS.}
\label{fig4}
\end{center}
\end{figure}

We now turn to addressing the chemical bonding through the ECOV criterion introduced above. In fig. \ref{fig3}b, along the y-axis, unitless  positive, negative and zero ECOV magnitudes point to bonding, anti-bonding and non-bonding interactions respectively. The bonding between the TM ions and C/N is mainly ensured by the itinerant-like electrons in the lower part of the valence band, i.e., in the energy range [-4,-2 eV] where one observes that the Cr-C bonding is twice larger than the V-N one. This confirms the finding of the ratio k($\frac{Cr-C}{V-N}$) from the vibrational spectra analysis of the isolated molecules providing the same trend without having the same value.  At the Fermi level there are mainly t$_{2g}$-like orbitals which show antibonding character;  but this is not the major issue of the bonding  within the system as these orbitals are mainly engaged in the magnetic instability as discussed above. In the conduction band above E$_F$ the antibionding counterparts of the Cr-C and V-N are found in the energy range [2, 6 eV]. 

Furthermore it can be suggested from the analyses above from the antibonding interaction between V and C as well Cr and N that V must be bonded to N and not C, contrary to Cr which must be bonded to C.  In order to provide rationalize this observation,  an additional calculation was carried out with the hypothesis of Cr-N and V-C direct bonds. The resulting total energy is found to destabilize the system by 1.32 eV. Thus the system is stable in the experimentally proposed configuration. This result is further confirmed using pseudo-potential calculations with the VASP code, where the destabilization amounts to 1.28 eV. Such energy criteria are features that can be used as a predictive tool for unknown systems.  

\subsubsection{Magnetic calculations for KV[Cr(CN)$_6$]}
From spin occupation for each atom species at self convergence, the total magnetization is close to 6 for the unit cell. This agrees  with Hund's rules' spin-only magnetic moments of 3 for both V and Cr. From the partial DOS cf. fig. \ref{fig4}a, the spin polarization clearly affects the two TM Cr and V. Also, we have a very low DOS at the E$_V$ pointing to the top of valence band. Total energy favors the ferromagnetic spin state state by 1.15 eV versus the non magnetic configuration.  From Table 3, all other atomic species, including the empty spheres exhibit negligible moments.
\subsubsection{Antiferromagnetic calculations}
Using a double cell with anti-parallel directions, the total magnetization is equal to zero and the system becomes an insulator with a gap opening of $\sim ~1 eV$ cf. fig. \ref{fig4}b. The spin magnetic moments for the ferro- FM and antiferromagnetic AFM configurations are given in table 3. Their magnitudes show a small lowering with respect to the FM case; this is probably due to the symmetry breaking.
Total energy favors the AFM state by 0.154 eV versus FM. This points to the antiferromagnetic ground state for the KV[Cr(CN)$_6$] complex. 
\newline
The  calculated J value in the solid state from the same equation is equal to -418 cm$^{-1}$ which is found  41\% larger than the molecular calculations (-296.5 cm$^{-1}$). Such differences could be expected due to the different account of the chemical system as an isolated molecule versus an extended solid; nevertheless the trends are the same.  

\subsubsection{Calculations for KNi[Cr(CN)$_6$]}
Following the same calculation protocol, the nonmagnetic configuration is found unstable to the ferromagnetic one by 1.6 eV. 
 From the magnetic calculations, the total magnetic moment of the cell is equal to 5 in agreement with Hund rule Ni$^{2+}$ and Cr$^{3+}$. The atomic magnetic moments are given in the table 5. Contrary to the case of KV[Cr(CN)$_6$] a hypothetic AFM configuration is found less stable than FM thus confirming the magnetic ground state for the system (see table 4). 

The energies of the two complexes in the three configurations NM, FM and AFM are given in table 5. \\
\newline

The difference in values between the $|J|$ of KNi[Cr(CN)$_6$] and KV[Cr(CN)$_6$](40.5 cm$^{-1}$ for 296.5) is that the factor 4$\beta$S in the relation of J given by the Kahn's model J = 2k + 4$\beta$S \cite{Girerd} is equal to 0 for the KNi[Cr(CN)$_6$] where it is negative in the second one. k is the bielectronic exchange integral (positive) between the two non orthogonalized magnetic orbitals a and b, $\beta$ is the corresponding monoelectronic resonance or transfer integral (negative) and S the monoelectronic overlap integral (positive) between a and b. It should be noted that 4$\beta$S is more larger than 2k which gave a biggest value of $|J|$ for the KV[Cr(CN)$_6$] complex. 

\section{Conclusion}
The goal of this work was to carry out a theoretical study of the exchange coupling constants in two Prussian blues analogues as well as the magnetic order and chemical bonding. We chose a compound with a strong antiferromagnetic coupling between the two TM, KV[Cr(CN)$_6]$ and with the same local spin, the other KNi[Cr(CN)$_6$] with a ferromagnetic coupling. The sign of  J, the values and the atomic spin densities obtained with the molecular calculations are in good agreement with those obtained by other authors. The complex with t$_{2g}^m$-t$_{2g}^n$ pairs give a  strong antiferromagnetic interaction (case of Cr-V) while t$_{2g}^m$-t$_{2g}^6$e$_g^2$ pairs yielding ferromagnetic coupling (case of Cr-Ni). This is an agreement with the conclusions given by Ruiz \cite{Ruiz}. We note that the T$_C$ values obtained from coupling constants in these two dinuclear models are significantly higher than the experimental ones.
\newline
In the extended solid state approach, we have applied the all electrons ASW method to describe the electronic and magnetic properties as well as the chemical bonding of the two compounds. Our calculations yield stable and insulating magnetic ground states for both complexes. From the ECOV allowing for a qualitative chemical bonding analysis between atomic species, it has clearly been established that Cr must be bonded to C and that V to N. The Cr-C bond strength is found larger than the V-N one; a result confirmed  by the vibrational spectra analysis. From the antiferromagnetic calculations, we proved that the KV[Cr(CN)$_6$] complex  is an insulating antiferromagnet while KNi[Cr(CN)$_6$] is an insulating ferromagnet.

The issue of the solid state calculations beside molecular ones is a major feature brought by this work by presenting the two approaches as complementary and bringing new results such as casting a quantitative description for the energy discrimination between different magnetic configurations and different chemical bonding possibilities through the isomerism.
\section{Acknowledgments}
Lara Kabalan thanks the CNRS$L$, Lebanese Council of Research, for her  Ph.D. scholarship.
We acknowledge computational facilities provided by the University Bordeaux 1 within the M3PEC ${Mesocentre~Regional}$ (http://www.m3pec.u-bordeaux1.fr) supercomputers. We also thank Dr. Laurent Duccasse (ISM, University Bordeaux 1) for providing access to the {\it JAGUAR} code.

\newpage
\begin{table}[htbp]
\begin{center}
\caption{Antiferromagnetic and ferromagnetic exchange interactions between d orbitals of Cr and M ions.} 
\vspace*{0.5 cm}
\begin{tabular}{ccc|c}
\hline
\hline
Cr & CN&M&Type of interaction \\
\hline
 e$_g$(empty)&p$_\sigma $&e$_g$(empty)&AFM\\
t$_{2g}$(half-filled)&p$_\pi $&t$_{2g}$(half-filled)&\\
\hline
e$_g$(empty)&p$_\sigma $&e$_g$(partially filled)&FM\\
t$_{2g}$(half-filled)&p$_\pi $&t$_{2g}$(filled)& \\
\hline
\hline
\end{tabular}
\label{}
\end{center}
\end{table}

\vspace*{1 cm} 
\begin{table}[htbp]
\begin{center}
\caption{Atomic spin densities on Cr, M for the two systems calculated with B3LYP functional and LanL2DZ basis set.} 
\vspace*{0.5 cm}
\begin{tabular}{c|ccc}
\hline
\hline
System&2S+1&Cr&M\\
\hline
 Cr$^{III}$-V$^{II}$&7&3.10&2.57\\
\hline
 Cr$^{III}$-V$^{II}$&1&2.98&-2.50\\
\hline
  Cr$^{III}$-Ni$^{II}$&6&3.10&1.65\\
\hline
  Cr$^{III}$-Ni$^{II}$&2&3.08&-1.65\\
\hline
\hline
\end{tabular}
\label{}
\end{center}
\end{table}

\vspace*{1 cm} 
\begin{table}[htbp]
\begin{center}
\caption{Spin magnetic moments of each atom of the cell for the KV[Cr(CN)$_6$] complex in the ferromagnetic (FM) and antiferromagnetic (AFM) configurations. ES1, ES2, ES3 are empty spheres entered within the ASA approximation (cf. text).}
\vspace*{1cm}
\begin{tabular}{c|cc}
\hline
&~~~FM&AFM\\
\hline

Atoms &~~~~~~~Magnetic moment value in $\mu_B$&\\
\hline

 K&0.005&-0.001\\
\hline
 V&2.409&-2.273\\
\hline
Cr&2.681&2.529\\
\hline
  C&-0.006&-0.076\\
\hline
 N&0.008&-0.045\\
 \hline
ES1&0.004&-0.041\\
\hline
ES2&0.028&-0.047\\
\hline
ES3&0.042&-0.066\\
\hline
\hline
cell&$\sim6$&$\sim0.0$\\
\hline
\hline
\end{tabular}
\label{}
\end{center}
\end{table}

\vspace*{1 cm} 
\begin{table}[htbp]
\begin{center}
\caption{Spin magnetic moments of each atom of the cell or the KNi[Cr(CN)$_6$] complex in the ferromagnetic (FM) and antiferromagnetic (AFM) configuartions. ES1, ES2, ES3 are empty spheres entered within the ASA approximation (cf. text).} 
\vspace*{0.5 cm}
\begin{tabular}{c|cc}
\hline
\hline
&~~~FM&AFM\\
\hline
Atoms &~~~~~~~Magnetic moment value in $\mu_B$&\\
\hline
 K&-0.000&-0000\\
\hline
 Ni&1.350&-1.352\\
\hline
Cr&2.720&2.715\\
\hline
  C&-0.043&-0.236\\
\hline
 N&0.134&-0.287\\
 \hline
ES1&0.00&-0.06\\
\hline
ES2&0.01&-0.07\\
\hline
ES3&0.03&-0.09\\
\hline
\hline
cell&$\sim 5$&$\sim 0.0$\\
\hline
\hline
\end{tabular}
\label{}
\end{center}
\end{table}

\vspace*{1 cm} 
\begin{table}[htbp]
\begin{center}
\caption{Energy difference relative to the nonmagnetic state of the ferromagnetic and antiferromagnetic configurations for the two complexes,  E$_{NM}$ = -85254.5512 eV for  KV[Cr(CN)$_6$] and  E$_{NM}$ = -100783.3848 eV for KNi[Cr(CN)$_6$]}
\vspace*{0.5 cm}
\begin{tabular}{c|cc}
\hline
\hline
$\Delta _E$(eV)&${\rm V[Cr(CN)_6]}$ & ${\rm Ni[Cr(CN)_6]}$\\
 \hline
 NM&0& 0\\
 \hline
 FM&-1.16&-1.59\\
 \hline
 AFM&-1.31&-1.51\\
\hline
\hline
\end{tabular}
\label{}
\vspace*{0.5 cm} 
\end{center}
\end{table}
\newpage

\end{document}